\documentclass{article}

\usepackage{microtype}
\usepackage{graphicx}
\usepackage{subfigure}
\usepackage{booktabs} 
\usepackage{amsthm}
\usepackage{amsmath,amsfonts,bm}
\usepackage[version=4]{mhchem}
\usepackage{siunitx}
\usepackage{amsthm}
\usepackage{amsmath,amsfonts,bm}

\makeatletter
\def\BState{\State\hskip-\ALG@thistlm}
\makeatother

\def\eqref#1{equation~\ref{#1}}

\def\1{\mathbb{I}}

\def\rs{{\textnormal{s}}}

\def\ry{{\textnormal{y}}}

\def\rx{{\textnormal{x}}}

\def\rvy{{\mathbf{y}}}

\def\rvx{{\mathbf{x}}}

\def\vy{{\bm{y}}}

\def\vx{{\bm{x}}}

\DeclareMathAlphabet{\mathsfit}{\encodingdefault}{\sfdefault}{m}{sl}
\SetMathAlphabet{\mathsfit}{bold}{\encodingdefault}{\sfdefault}{bx}{n}

\usepackage{siunitx}
\usepackage{comment}
\usepackage{hyperref}

\usepackage[accepted]{icml2021}

\icmltitlerunning{} 

\begin{document}

\twocolumn[
\icmltitle{OCDE: Odds Conditional Density Estimator}



\icmlsetsymbol{equal}{*}

\begin{icmlauthorlist}
\icmlauthor{Alex Akira Okuno}{equal,usp}
\icmlauthor{Felipe Maia Polo}{equal,mich,ai}
\end{icmlauthorlist}

\icmlaffiliation{usp}{Department of Statistics, Institute of Mathematics and Statistics, University of São Paulo, Brazil}
\icmlaffiliation{mich}{Department of Statistics, University of Michigan, USA}
\icmlaffiliation{ai}{Advanced Institute for Artificial Intelligence (AI2), Brazil}

\icmlcorrespondingauthor{Alex Akira Okuno}{akira.okuno@outlook.com}
\icmlcorrespondingauthor{Felipe Maia Polo}{felipemaiapolo@gmail.com}

\icmlkeywords{Conditional Density Estimation, CDE, Density Ratio Estimation, Probabilistic Classifier, Nonparametric Statistics, Statistics, Machine Learning}

\vskip 0.3in
]



\printAffiliationsAndNotice{\icmlEqualContribution} 

\begin{abstract}
Conditional density estimation (CDE) models can be useful for many statistical applications, especially because the full conditional density is estimated instead of traditional regression point estimates, revealing more information about the uncertainty of the random variable of interest. In this paper, we propose a new methodology called Odds Conditional Density Estimator (OCDE) to estimate conditional densities in a supervised learning scheme. The main idea is that it is very difficult to estimate $p_{\rvx,\rvy}$ and $p_{\rvx}$ in order to estimate the conditional density $p_{\rvy|\rvx}$, but by introducing an instrumental distribution, we transform the CDE problem into a problem of odds estimation, or similarly, training a binary probabilistic classifier. We demonstrate how OCDE works using simulated data and then test its performance against other known state-of-the-art CDE methods in real data. Overall, OCDE is competitive compared with these methods in real datasets.
\end{abstract}

\section{Introduction}

 Conditional density estimation (CDE) consists of estimating the conditional density of a random vector $\rvy\in\mathbb{R}^k$ given a random vector $\rvx\in\mathbb{R}^d$, where generally $k=1$. It
 is a difficult problem in the sense that the full conditional distribution $p_{\rvy|\rvx}$ has to be estimated for many values of $\rvx$. In this context, the CDE problem is a generalization of the regression problem, where instead of estimating only the expectation $\mathbb{E}[\rvy|\rvx]$, we model the full conditional distribution, which gives much more information about the uncertainty of the random variable of interest.
 
 In many statistical applications, though, in particular when the conditional distribution $p_{\rvy|\rvx}$ is well-behaved, i.e. uni-modal and symmetric, $\mathbb{E}[\rvy|\rvx]$ is generally informative enough. But, this density is often asymmetric and displays multi-modality, in which case the full conditional density would be much more informative. Besides, having a full density allows us to calculate a diversity of statistical quantities like moments including higher order ones, probability intervals, quantiles, non-trivial expectations and so on. 
 
 In this paper, we propose a novel methodology called Odds Conditional Density Estimator (OCDE) in order to address the CDE problem. The idea behind OCDE is that we can transform the problem of conditional density estimation into a classification problem that can be solved by training a binary probabilistic classifier, i.e., neural networks or gradient boosting algorithms.

\section{Related Work}

Regarding the problem of estimating a conditional density $p_{\rvy|\rvx}$ in an i.i.d. context, there have been several previous methodologies in the literature. At first, the CDE was introduced by \citet{Rosenblatt}, whose approach was to estimate $p_{\rvx,\rvy}$ and $p_{\rvx}$ with kernel density estimators, and then calculating their ratio. However, kernel methods for density estimation, in general, do not scale well with the dimension of covariates and sample size. Several other papers have developed upon this problem with other approaches such as polynomial regressions \citep{FanYaoTon96}, least squares \citep{sugiyama2010least} and quantile estimation \citep{Takeuchi2}.

A recent CDE method that overcomes the aforementioned issues is FlexCoDE \citep{IzbickiLeeFlexCode}. This is due to the fact that this method can transform the CDE problem into a regression problem, making the conditional density estimator to inherit the properties of the regression method used (for example, variable selection, regularization and good properties in high-dimensional settings). Our work relates to FlexCoDE in the sense that our method also transforms the CDE problem into a supervised model training problem, but in our case, we need to solve a classification problem.

Another interesting CDE method is NN-KCDE (or nearest-neighbors kernel CDE) \citep{izbicki2018abc}. This method is the usual kernel density estimator using only the closest points, in covariate space, to the target point $\rvx$ \citep{izbicki2018abc}. It is possible to tune the number of neighbors hyperparameter. 

Our method relies on directly estimating the density ratio $\frac{p_{\rvx,\rvy}}{p_{\rvx}}$ using a method called “Probabilistic Classification" or sometimes referred as “Logistic Regression based method" \cite{sugiyama2012machine, sugiyama2012density}. Interestingly, \citet{sugiyama2010least} states that “Logistic Regression based method" may not be employed for CDE because $p_{\rvx,\rvy}$ and $p_{\rvx}$ do not share the same domains. We get around this problem by introducing an instrumental random variable, which makes the domains match. Given that, our method is an extension of a method for density estimation presented in Section 14.2.4 of \citet{hastie2009elements}. For the best of our knowledge, a method with this foundation has not been used for conditional density estimation yet. 
\section{Odds Conditional Density Estimator (OCDE)}

In this section, we present the OCDE for multivariate target vector $\rvy$ and feature vectors $\rvx$. 
Suppose we have a set of data points $\mathcal{D}=\{(\vx_i,\vy_i)\}_{i=1}^n$ sampled independently from $P_{\rvx,\rvy}$ with probability density function (p.d.f.) $p_{\rvx,\rvy}$. Our objective is to estimate $p_{\rvy|\rvx}$ using the dataset $~\mathcal{D}$. Suppose that $\rvy$ is a random vector that assumes values in $\text{support}(p_\rvy) \subseteq \mathbb{R}^k$. In order to continue, let us introduce an
instrumental/artificial random variable $\tilde{\rvy}$ with known distribution $P_{\tilde{\rvy}}$, independent from $(\rvx,\rvy)$, with known p.d.f. $p_{\tilde{\rvy}}$ such that $\text{support}(p_\rvy)\subseteq\text{support}({p_{\tilde{\rvy}}})$. 

See that we can rewrite $p_{\rvy|\rvx}$ as follows:
\begin{align}
  p_{\rvy|\rvx}(\vy|\vx)&=\frac{p_{\rvx,\rvy}(\vx,\vy)}{p_{\rvx}(\vx)} \\[.5em]
  &={\frac{p_{\tilde{\rvy}}(\vy)}{p_{\tilde{\rvy}}(\vy)}}\frac{p_{\rvx,\rvy}(\vx,\vy)}{p_{\rvx}(\vx)} \\[.5em]
  &={p_{\tilde{\rvy}}(\vy)}\frac{p_{\rvx,\rvy}(\vx,\vy)}{p_{\rvx,\tilde{\rvy}}(\vx,\vy)}
\end{align}

Our objective of estimating $p_{\rvy|\rvx}$ can be reduced in estimating the density ratio $\frac{p_{\rvx,\rvy}}{p_{\rvx,\tilde{\rvy}}}$. To that end, we adopt the “Probabilistic Classification" method for density ratio estimation \cite{sugiyama2012density}, which is detailed next. Consider a random vector $(\rvx',\rvy')\sim P_{\rvx',\rvy'}$ with p.d.f. $p_{\rvx',\rvy'}$ and another artificial random variable $\rs \sim \textup{Bernoulli}(1/2)$, given that
\begin{align}
&p_{\rvx,\rvy}(\vx,\vy)=p_{\rvx',\rvy'|\rs}(\vx,\vy|\rs=1)\\[.5em]
&p_{\rvx,\tilde{\rvy}}(\vx,\vy)=p_{\rvx',\rvy'|\rs}(\vx,\vy|\rs=0)
\end{align}

That is, $\rs$ is an indicator variable that tells us if data point comes from $P_{\rvx,\rvy}$ or $P_{\rvx,\tilde{\rvy}}$. From Bayes rule, it follows that:
\begin{align}
  p_{\rvy|\rvx}(\vy|\vx)&={p_{\tilde{\rvy}}}(\vy)\frac{p_{\rvx',\rvy'|\rs}(\vx,\vy|\rs=1)}{p_{\rvx',\rvy'|\rs}(\vx,\vy|\rs=0)} \\[.5em]
  &= {p_{\tilde{\rvy}}}(\vy) \frac{p_{\rs|\rvx',\rvy'}(\rs=1|\vx,\vy)}{p_{\rs|\rvx',\rvy'}(\rs=0|\vx,\vy)}
\end{align}

From our original dataset $\mathcal{D}$, we derive another dataset $\tilde{\mathcal{D}}=\{(\vx_i,\tilde{\vy}_i)\}_{i=1}^{n}$, with values $\{\tilde{\vy}_i\}_{i=1}^{n}$ being instances of $\tilde{\rvy}$. Then, we create artificial labels for the data points of $\mathcal{D}$ and $\tilde{\mathcal{D}}$, where the first samples receive labels 1 and the second receive labels 0. In other words, we label samples according to the variable $\rs$. Then, we train a \textit{non-linear} binary probabilistic classifier $\widehat{p}_{\rs|\rvx',\rvy'}$ discriminating samples from $\mathcal{D}$ and $\tilde{\mathcal{D}}$. 

Finally, our estimator for the conditional density, OCDE, is given by
\begin{align}
  \widehat{p}_{\rvy|\rvx}(\vy|\vx)&={p_{\tilde{\rvy}}}(\vy)\frac{\widehat{p}_{\rs|\rvx',\rvy'}(\rs=1|\vx,\vy)}{\widehat{p}_{\rs|\rvx',\rvy'}(\rs=0|\vx,\vy)} 
\end{align}

Some useful information about OCDE are the following:
\begin{itemize}
    \item Given that $\widehat{p}_{\rs|\rvx',\rvy'}$ should be a good estimate for the true conditional distribution $p_{\rs|\rvx',\rvy'}$, we advise practitioners to optimize the binary cross-entropy/log loss when training or choosing the classifier's hyperparameters. Depending on the model adopted, probability calibration might be necessary;
    \item It can be the case that $\widehat{p}_{\rvy|\rvx}(\vy|\vx)$ does not integrate to 1 for all possible values of $\vx$. To fix this problem, we can discretize $\widehat{p}_{\rvy|\rvx}(\vy|\vx)$ into a histogram\footnote{Histograms of $100$ or $1000$ bins, for example.} and then normalize the histogram itself. In practice, we always adopt this strategy;
    \item The choice of $p_{\tilde{\rvy}}$ is non-trivial. A standard choice is to assume that $\tilde{\rvy}$ is uniformly distributed in some reasonable bounded subset of $\mathbb{R}^k$ that can be chosen according to the training set samples;
    \item In theory, the instrumental random vector $\tilde{\rvy}$ can be dependent on $\rvx$, but we do not explore this scenario in this paper.
\end{itemize}


\section{Experiments}

\subsection{Toy Experiment}

\begin{figure}[h!]
    \centering
    \includegraphics[width=0.4\textwidth]{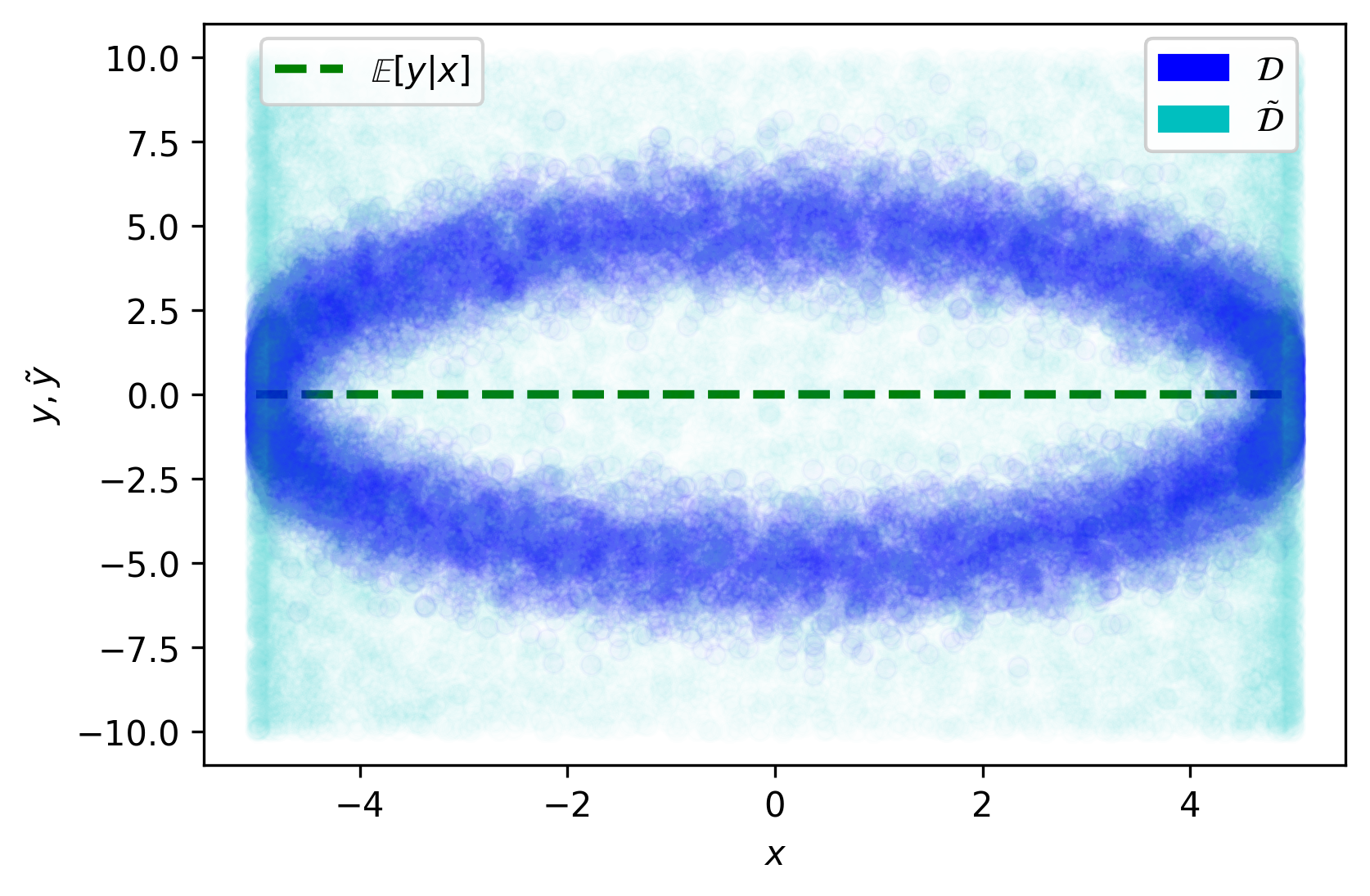}
    \caption{Real ($\mathcal{D}$) and artificial ($\tilde{\mathcal{D}}$) datasets used to estimate $p_{\ry|\rx}$. We also plot $\mathbb{E}[\ry|\rx]$ to show it maps to points in the domain of $\ry$ that, in most cases, have no probability density.}
    \label{fig:dists}
\end{figure}

In this experiment, we sample from $P_{\rx,\ry}$ indirectly. If $\theta_i \sim U[0,2\pi]$ and $\epsilon_i \sim N(0,1)$, we use the following functional forms to sample $\rx_i$ and $\ry_i$, for $i ={1,...,n}$: $\rx_i=5 \cos (\theta_i)$ and $\ry_i=5 \sin (\theta_i) + \epsilon_i$. Given that we sample $(\theta_i, \epsilon_i)$ independently from $(\theta_j, \epsilon_j)$, if $i \neq j$, then $(\rx_i, \ry_i)$ is also independent from  $(\rx_j, \ry_j)$. In this example, we assume the instrumental random variables $\{\tilde{\ry}_i\}_{i=1}^n$ are sample independently from $U[-10,10]$, hence having a known p.d.f. In the following, we create the datasets $\mathcal{D}$ and $\tilde{\mathcal{D}}$, with $n=10000$, and plot them in Figure \ref{fig:dists}.

We can make two important observations about Figure \ref{fig:dists}: (i) for certain values of $\rx \in (-5,5)$, the conditional density $p_{\ry|\rx}$ is multi-modal and (ii) the expectation $\mathbb{E}[\ry|\rx]$ maps to points in the domain of $\ry$ that, in most cases, actually have no probability density. 

\begin{figure}[h!]
    \centering
    \includegraphics[width=0.4\textwidth]{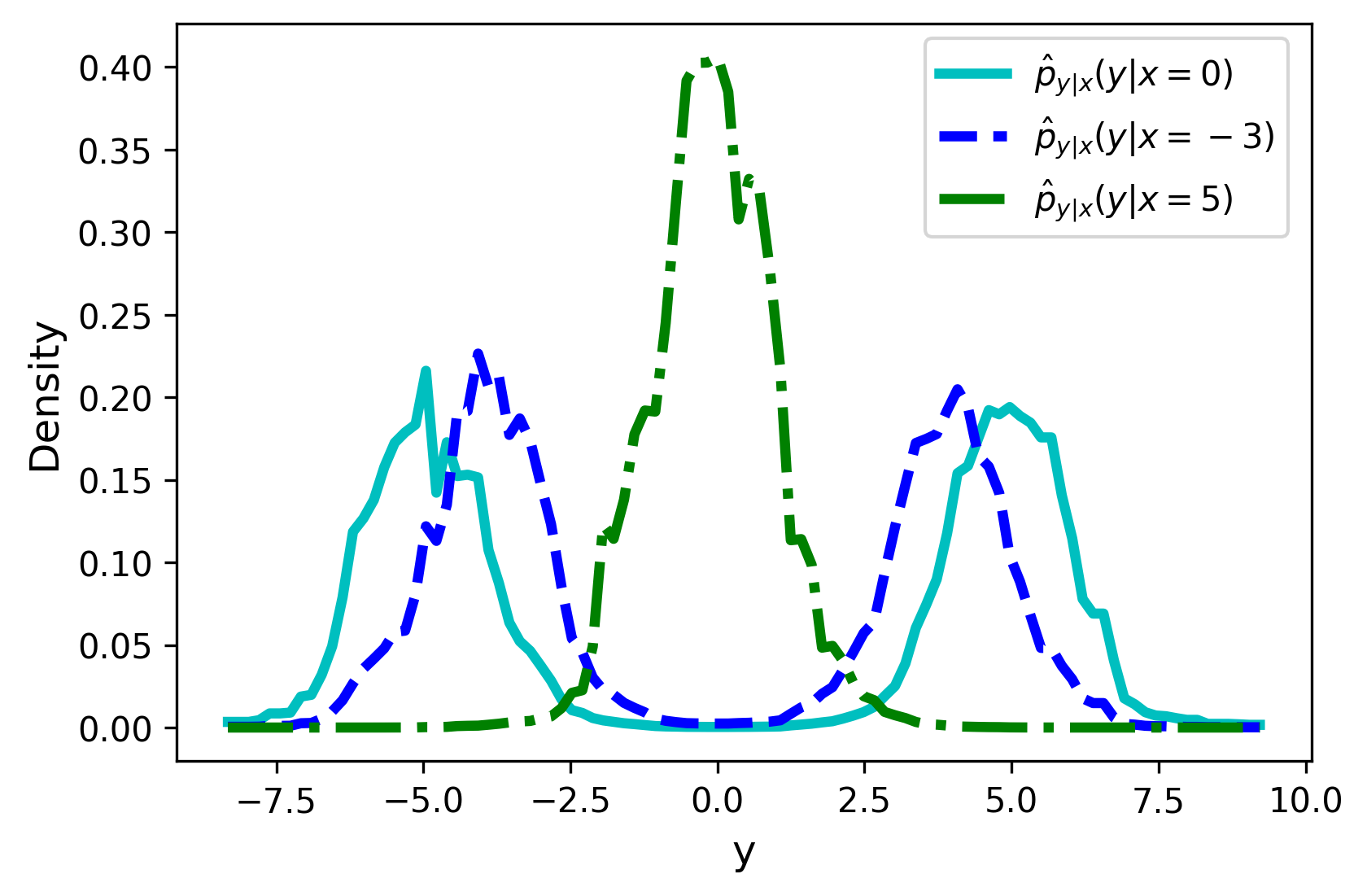}
    \caption{OCDE estimates for the conditional distribution of $\ry$ while varying the values of $\rx$.}
    \label{fig:cdes1}
\end{figure}

In this experiment, we use the CatBoost Classifier\footnote{See \url{https://catboost.ai}.} \cite{prokhorenkova2017catboost} with default hyperparameters as our binary probabilistic classifier. Figure \ref{fig:cdes1} lets us see some examples of OCDE estimates for the conditional distribution while varying the values of $\rx$. The distinction in the modality of the conditional distribution is very clear when we compare $\widehat{p}_{\ry|\rx}(y|\rx=0)$ against $\widehat{p}_{\ry|\rx}(y|\rx=5)$. We do not intend to evaluate our model with this toy experiment, but we have a nice visual intuition that our model incorporates the change of modality reasonably well in a simple setup.

Given that we use the CatBoost Classifier as our binary probabilistic classifier, it is expected that the conditional density $\widehat{p}_{\ry|\rx}$ is a non-smooth function, which is clear in Figure \ref{fig:cdes1}. It is possible to approximate $\widehat{p}_{\ry|\rx}$ using the Fast Fourier Transform (FFT) algorithm for a smoother estimate, for example. 



\subsection{Real Datasets Experiments}

\begin{table*}[ht] 
 \centering 
 \caption{CDE Loss ($\pm$ std. error) obtained by the four alternatives for conditional density estimation in the \textit{Raw} and \textit{Debiased} approaches. In summary, the results presented in this table puts OCDE among the state-of-the-art methods for conditional density estimation.} 
 \medskip
 \label{tab:performance}%
 \resizebox{480pt}{!}{ 
 \setlength\tabcolsep{1.25pt}\begin{tabular}{c|rrrr|rrrr} 
 \hline 
 \multicolumn{1}{c}{} &  \multicolumn{4}{c}{Raw} & \multicolumn{4}{c}{Debiased}  \\ 
 \hline 
 Dataset & OCDE~~~~ & OCDE (Smooth) & FlexCoDE~  & NN-K~~~~~~ & OCDE~~~~ & OCDE (Smooth) & FlexCoDE~  & NN-K~~~~~~ \\ 
 \hline 
abalone 
& $ 67.93 \pm  1.15 $ & $ -1.00  \pm  0.00 $ & $ -4.07  \pm 0.12 $ & $ \mathbf{-8.81  \pm  0.25} $ & $ -4.81  \pm  0.03 $ & $ -4.81  \pm 0.03 $ & $ -4.20  \pm  0.06 $ & $ -4.92  \pm  0.05 $ \\ 
ailerons 
& $ 39.14 \pm  0.80 $ & $ -1.12  \pm  0.00 $ & $ -6.40  \pm 0.15 $ & $ -7.81  \pm  0.21 $ & $ \mathbf{-8.07  \pm  0.03} $ & $ \mathbf{-8.08  \pm 0.03} $ & $ -7.44  \pm  0.06 $ & $ \mathbf{-8.08  \pm  0.06} $ \\ 
bank32nh 
& $ \mathbf{-25.14 \pm  1.19} $ & $ -21.89  \pm  1.03 $ & $ \mathbf{-26.72  \pm 1.39} $ & $ -11.69  \pm  0.42 $ & $ -5.21  \pm  0.22 $ & $ -5.22  \pm 0.23 $ & $ -4.96  \pm  0.19 $ & $ -4.29  \pm  0.28 $ \\ 
bank8FM 
& $ \mathbf{-10.54 \pm  0.61} $ & $ -9.73  \pm  0.51 $ & $ -9.44  \pm 0.45 $ & $ -3.08  \pm  0.19 $ & $ \mathbf{-10.63  \pm  0.30} $ & $ \mathbf{-10.63  \pm 0.42} $ & $ -8.97  \pm  0.32 $ & $ -9.03  \pm  0.32 $ \\ 
cal housing 
& $ \mathbf{-6.03 \pm  0.36} $ & $ \mathbf{-5.77  \pm  0.37} $ & $ \mathbf{-5.56  \pm 0.18} $ & $ -4.61  \pm  0.18 $ & $ -5.20  \pm  0.04 $ & $ -5.21  \pm 0.04 $ & $ -5.34  \pm  0.08 $ & $ -5.25  \pm  0.06 $ \\ 
cpu act 
& $ \mathbf{-15.78 \pm  0.79} $ & $ -14.05  \pm  0.58 $ & $ -14.53  \pm 0.39 $ & $ -14.37  \pm  0.56 $ & $ -13.13  \pm  0.47 $ & $ -13.14  \pm 0.42 $ & $ -13.38  \pm  0.30 $ & $ \mathbf{-14.85  \pm  0.31} $ \\ 
cpu small 
& $ \mathbf{-14.98 \pm  0.75} $ & $ -13.13  \pm  0.65 $ & $ -13.18  \pm 0.34 $ & $ -13.37  \pm  0.41 $ & $ -12.50  \pm  0.31 $ & $ -12.52  \pm 0.32 $ & $ -12.13  \pm  0.28 $ & $ -13.28  \pm  0.29 $ \\ 
delta ailerons 
& $ -5.48 \pm  0.24 $ & $ -5.69  \pm  0.20 $ & $ -9.21  \pm 0.23 $ & $ \mathbf{-13.03  \pm  0.37} $ & $ -10.69  \pm  0.06 $ & $ -10.70  \pm 0.06 $ & $ -11.17  \pm  0.13 $ & $ -12.01  \pm  0.12 $ \\ 
elevators 
& $ 0.82 \pm  0.83 $ & $ -6.79  \pm  0.24 $ & $ -8.30  \pm 0.19 $ & $ -8.45  \pm  0.32 $ & $ -8.89  \pm  0.06 $ & $ -8.90  \pm 0.05 $ & $ -8.30  \pm  0.12 $ & $ \mathbf{-9.59  \pm  0.06} $ \\ 
fried delve 
& $ -4.00 \pm  0.06 $ & $ -4.00  \pm  0.08 $ & $ -4.19  \pm 0.05 $ & $ -3.30  \pm  0.06 $ & $ \mathbf{-8.75  \pm  0.01} $ & $ \mathbf{-8.75  \pm 0.01} $ & $ -7.82  \pm  0.09 $ & $ -8.66  \pm  0.02 $ \\ 
puma32H 
& $ -3.62 \pm  0.04 $ & $ -3.62  \pm  0.05 $ & $ -4.96  \pm 0.07 $ & $ -1.79  \pm  0.05 $ & $ -5.91  \pm  0.03 $ & $ -5.91  \pm 0.03 $ & $ \mathbf{-6.16  \pm  0.07} $ & $ -5.80  \pm  0.04 $ \\ 
puma8NH 
& $ -2.28 \pm  0.04 $ & $ -2.28  \pm  0.05 $ & $ -1.89  \pm 0.06 $ & $ -2.06  \pm  0.06 $ & $ \mathbf{-2.32  \pm  0.03} $ & $ \mathbf{-2.32  \pm 0.04} $ & $ -2.12  \pm  0.04 $ & $ \mathbf{-2.37  \pm  0.03} $ \\ 
winequality 
& $ -2.79 \pm  1.78 $ & $ -13.18  \pm  1.01 $ & $ \mathbf{-26.78  \pm 0.71} $ & $ -19.25  \pm  0.47 $ & $ -2.50  \pm  0.03 $ & $ -2.50  \pm 0.03 $ & $ -14.30  \pm  0.24 $ & $ -2.92  \pm  0.07 $ \\ 
\hline 
\end{tabular}%
} 
\end{table*}%

For the following experiments, 13 regression datasets with no missing values have been selected from two different sources\footnote{\url{www.dcc.fc.up.pt/~ltorgo/Regression/DataSets.html} and \url{https://archive.ics.uci.edu/ml/datasets.php}}. In these datasets, the target variable is univariate while the number of features ranges from 6 to 40. For each one of the 13 datasets, we repeated the following pre-processing steps: (i) we kept up to 5,000 data points per dataset, (ii) normalized each column in every dataset, stretching values in the interval $[0,1]$, and (iii) randomly splitted the data points in a training set ($80\%$) and a test set ($20\%$). When necessary, we use part of the training set ($20\%$) as a validation set.

For this series of experiments, we take two approaches on estimating conditional density. In the first one (\textit{Raw}), we directly estimate the conditional density of $\ry$. In the second one (\textit{Debiased}), we first fit a regressor $\hat{f}(\vx)=\widehat{\mathbb{E}}[\ry|\rvx=\vx]$ using the training set and then estimate the conditional density of the residuals $\hat{\varepsilon} = \ry - \hat{f}(\rvx)$ given $\rvx$. In these experiments, the regressor $\hat{f}$ is a CatBoost Regressor \cite{prokhorenkova2017catboost}.

In each of the two approaches, we compare four alternatives for conditional density estimation: (i) OCDE, (ii) OCDE (Smooth), (iii) FlexCoDE, and (iv) NN-K. Each algorithm has the objective to minimize the CDE Loss \cite{IzbickiLeeFlexCode} on unseen data. More details on each one:
\begin{itemize}
    \item OCDE: We choose the CatBoost Classifier trained with default hyperparameters and early stopping rounds equals to 50 as our binary probabilistic classifier. The instrumental random variable $\tilde{\ry}$ is uniformly distributed in the interval bounded by the minimum and maximum values of the target variable in the training set;
    \item OCDE (Smooth): This estimator is almost identical to the OCDE. It only differs from the fact that we approximate $\widehat{p}_{\ry|\rx}$ using the Fast Fourier Transform (FFT) algorithm. The number of Fourier components minimizes the CDE Loss on the training set;
    \item FlexCoDE: We choose the XGBoost method as the regression method to train FlexCoDE, which was trained with default hyperparameters. Regarding the FlexCoDE itself, we chose $50$ as the max basis parameter and used the tuning procedure provided by its Python implementation;
    \item NN-K: In this estimator, we performed an extensive hyperparameter optimization, specifically for the number of neighbors and bandwidth level.
\end{itemize}

Table \ref{tab:performance} compares the CDE Loss ($\pm$ std. error), estimated on the test set, obtained by the four alternatives for conditional density estimation in the \textit{Raw} and \textit{Debiased} approaches. In summary, the results presented in Table \ref{tab:performance} puts OCDE among the state-of-the-art methods for conditional density estimation. Considering that two methods have the same performance if their error bars intercept, it is possible to see that OCDE has the best results in 8 datasets, while NN-K has the best results in 6 datasets, OCDE (Smooth) is the best in 5 datasets, and FlexCoDE is the best in 4 of them.

The average running times $\pm$ std. deviation (in seconds) for the four methods are the following: (i) OCDE: $2.26 \pm 1.97$; (ii) OCDE (Smooth): $128.96 \pm 47.59$; (iii) FlexCoDE: $125.18 \pm 76.5$; (iv) NN-K: $27.36 \pm 2.66$. Clearly, OCDE is the fastest one because it does not have any hyperparameter tuning phase. On the other hand, OCDE (Smooth) tunes the number of Fourier components, taking a longer time to run.


\section{Source code and OCDE Python package}

We are currently working to make our code and Python package available as soon as possible. The next version of this work will contain links to GitHub repositories containing it.

\section{Conclusion and Future Work}

In this paper, we propose a novel methodology called Odds Conditional Density Estimator (OCDE) in order to address the CDE problem. OCDE performed well against competitors in real dataset experiments. Future directions for this work could be: (i) testing how OCDE performs with different sample sizes and target dimensions, (ii) optimizing the choice of the instrumental random vector $\hat{\rvy}$, or (iii) testing OCDE against other benchmark approaches.

\nocite{langley00}

\bibliography{example_paper}
\bibliographystyle{icml2021}


\end{document}